\documentclass{iau} 
\usepackage{graphicx}
\usepackage{natbib}

\def\apj{\textit{ApJ}}
\def\apjl{\textit{ApJL}}

\def\aap{\textit{A\&A}}

\def\mnras{\textit{MNRAS}}

\def\nat{\textit{Nature }}
\def\na{\textit{New Astron.}}
\def\jcap{\textit{JCAP}}
\def\prd{\textit{Phys. Rev. D}}

\def\beq#1{\begin{equation}\label{#1}}
\def\eeq{\end{equation}}
\def\beqa#1{\begin{eqnarray}\label{#1}}
\def\eeqa{\end{eqnarray}}

\def\mycomment#1{\relax}

\title[LIGO binary BH mergers progenitors] 
{Progenitors of binary black hole mergers detected by LIGO}

\author[Konstantin Postnov \& Alexander Kuranov]   
{Konstantin Postnov 
 \and Alexander Kuranov}

\affiliation{Sternberg Astronomical Institute, Moscow M.V.Lomonosov State University \\
13, Universitetskij pr., 119234 Moscow, Russia
\\ email: {\tt pk@sai.msu.ru} }

\pubyear{2017}
\volume{329}  
\jname{The lives and death-throes of massive stars}
\editors{J.J. Eldridge, ed.}
\begin{document}

\maketitle

\begin{abstract}
Possible formation mechanisms of massive close binary black holes that can merge in the Hubble time to produce powerful gravitational wave bursts 
detected during advanced LIGO O1 science run are briefly discussed. The pathways include the evolution from field low-metallicity massive binaries, 
the dynamical formation in globular clusters and primordial black holes. Low effective black hole spins inferred for LIGO GW150914 and LTV151012 
events are discussed. Population synthesis calculations of the expected spin and chirp mass distributions from the standard field massive binary formation channel are presented for different metallicities (from zero-metal Population III stars up to solar metal abundance). We conclude that that merging binary black holes can contain systems from different formation channels, discrimination between which can be made with increasing
statistics of mass and spin measurements from ongoing and future gravitational wave observations.

\keywords{black hole physics, stars: evolution, (stars:) binaries (including multiple): close}
\end{abstract}

\firstsection 
\section{Introduction and historical remarks}
The epochal discovery of the first gravitational wave source GW150914 from coalescing binary black hole (BH) system 
\citep{LIGO-PRL} not only heralded the beginning of gravitational wave astronomy era, but also stimulated a wealth of works on 
fundamental physical and astrophysical aspects of the formation and evolution of binary BHs. The LIGO detection of  GW150914 
and of the second robust binary BH merging event GW151226 \citep{2016PhRvL.116x1103A} enables BH masses and spins before the merging, 
the luminosity distance to the sources and the binary BH merging rate in the Universe to be estimated \citep{2016PhRvX...6d1015A}. 
Astrophysical implications of these measurements were discussed, e.g., in \cite{2016ApJ...833L...1A,2016ApJ...818L..22A}. 

This discovery of gravitational waves from coalescing binary BHs was long awaited. 
Evolution of massive binary systems was elaborated in the 1970s to explain a rich variety of newly discovered galactic 
X-ray binaries \citep{1972NPhS..239...67V,1973NInfo..27....3T}. Formation of two relativistic compact remnants (neutron stars (NSs) or black holes) naturally
followed from the binary evolution scenario \citep{1973NInfo..27...70T,1975A&A....39...61F}. 
At the dawn of the LIGO Project, Tutukov and Yungelson 
\citep{1993MNRAS.260..675T} calculated, using the standard assumptions of massive binary evolution, the 
expected galactic merging rate of binary NSs and BHs. They pointed out that although the galactic merging rate of 
binary NSs is much larger than that of binary BHs, their detection rates by gravitational-wave interferometers 
can be comparable due to the strong 
dependence of the characteristic GW amplitude $h_c$ on the total mass $M=M_1+M_2$ of the coalescing binaries, $h_c\sim M^{5/2}$. 
A few years later, independent population synthesis calculations by the Scenario Machine code were reported in a series of papers 
\citep{1997NewA....2...43L,1997MNRAS.288..245L,1997AstL...23..492L}. They showed that in a wide range of possible BH formation parameters 
(masses,kick velocities) and under standard assumptions of the massive star evolution, the detection rate of binary BH mergings
should be much higher than that of binary NSs, and the first LIGO event should most likely to be a binary BH merging. Interestingly, 
the mean BH masses known at that time from dynamical measurements in galactic BH X-ray binaries were about 
10 $M_\odot$, which forced (cautiously) the authors of 
\citep{1997AstL...23..492L} to fix the parameter $k_{BH}=M_{BH}/M_c$, where $M_c$ is the mass of the star before the collapse,
around $\sim 0.3$ (see Fig. 4 in that paper) in order to produce the chirp mass of coalescing binary BHs around $15 M_\odot$. Taking $k_{BH}=1$, one immediately obtains the BH masses around 30-40 $M_\odot$, which seemed outrageously high at that time. 

Starting from the end of the 1990s, various groups have used different population synthesis codes to calculate the merging rates of double compact objects 
(see especially many papers by the Polish group based on the StarTrack code \citep{2002ApJ...572..407B,2012ApJ...759...52D}), yielding a wide range of possible 
BH-BH merging rates (see e.g. Table 6  in \cite{2014LRR....17....3P}). Clearly, the degeneracy of binary evolution and BH formation parameters
has been so high \citep{2010CQGra..27q3001A} that only real observations could narrow the wide parameter range.


\section{Standard scenario of binary BH formation}

The standard scenario of double BH formation from field stars is based on well-recognized evolution of single massive stars \cite{2002RvMP...74.1015W}. 
To produce a massive BH with $M\simeq 10M_\odot$ in the end of evolution, the progenitor star should have a large mass and low mass-loss rate. 
The mass-loss rate is strongly dependent on the metallicity, which plays the key role in determining the final mass of 
stellar remnant (see \cite{2015MNRAS.451.4086S} and N. Yusof's contribution in this conference). The metallicity effects were included in the population synthesis calculations \citep{2013ApJ...779...72D}, and the most massive BHs were found to be produced by the low-metallicity progenitors.
Here early metal-free Population III stars provide an extreme example, see calculations by \cite{Kinugawa2014,2016MNRAS.460L..74H}. After the discovery of GW150914, 
several independent population synthesis calculations were performed to explain the observed masses of binary BH in GW150914 and the inferred binary BH 
merging rate $\sim 9-240$~Gpc$^{-3}$yr$^{-1}$ \citep{2016PhRvX...6d1015A} \citep[see, among others, e.g.][]{2016Natur.534..512B,2016MNRAS.462.3302E,2017NewA...51..122L}. 

In addition to the metallicity that affects the intrinsic evolution of the binary components, 
the most important uncertainty in the binary evolution is the efficiency of the common envelope (CE) stage which 
is required to form a compact double BH binary merging in the Hubble time. The common envelope stage remains a highly debatable issue. 
For example, in recent hydro simulations \citep{2016ApJ...816L...9O} a low CE efficiency was found, while successful CE calculations 
were reported by other groups (see, e.g., N. Ivanova contribution at this conference).  
Another recent study \citep{2017MNRAS.465.2092P} argues that 
it is possible to reconcile the BH formation rate through the CE channel taking into account the stability of mass transfer in massive binaries 
in the Hertzsprung gap stage, which 
drastically reduces the otherwise predicted overproduction of binary BH merging rate in some population synthesis calculations.
Also, the so-called stable 'isotropic re-emission' mass transfer mode can be realized in high-mass X-ray binaries with massive BHs, thus helping to avoid 
the merging of the binary system components in the common envelope \citep{2017arXiv170102355V}. This stable mass transfer mode 
can explain the surprising stability of kinematic characteristics observed in the galactic microquasar SS433 \citep{2008ARep...52..487D}.

Of course, much more
empirical constraints on and hydro simulations of the common evolution formation and properties are required, but the formation 
channel with common envelope of binary BHs with properties similar to GW150914  remains quite plausible.

\section{Other scenarios}

To avoid the ill-understood common envelope stage, several alternative scenarios of binary BH formation from massive stars were proposed. 
For example, in short-period  massive binary systems chemically homogeneous evolution due to rotational mixing can be realized. The stars remain
compact until the core collapse, and close binary BH system is formed without common envelope stage \citep{2016MNRAS.458.2634M,2016MNRAS.460.3545D,2016A&A...588A..50M}.
In this scenario, a pair of nearly equal massive BHs can be formed with the merging rate comparable to the empirically inferred from the first LIGO observations. 

Another possible way to form massive binary BH system is through dynamical interactions in a dense stellar systems (globular clusters). This 
scenario was earlier considered by \cite{1993Natur.364..423S}. In the core of a dense globular clusters, stellar-mass BH form multiple systems, and 
BH binaries are dynamically ejected from the cluster. 
This mechanism was shown to be quite efficient in producing 30+30 $M_\odot$ merging binary BHs \citep{2016ApJ...824L...8R}, and 
binary BH formed in this way can provide a substantial fraction of all binary BH mergings in the local Universe \citep{2016PhRvD..93h4029R}. 

Finally, there can be more exotic channels of binary BH formation. For example, primordial black holes (PBHs) 
formed in the early Universe can form pairs which could be 
efficient sources of gravitational waves \citep{1997ApJ...487L.139N}. After the discovery of GW150914, the interest to binary PBHs
has renewed \citep{2016PhRvL.116t1301B}. Stellar-mass PBHs can form a substantial part of dark matter in the Universe \citep{2016PhRvD..94h3504C}. The PBHs formed
at the radiation-dominated stage can form pairs like GW150914 with the merging rate compatible with empirical LIGO results, being only a small fraction of all 
dark matter \citep{2016arXiv160404932E,2016PhRvL.117f1101S}. Different class of PBHs with a universal log-normal mass spectrum produced in the frame of a modified Affleck-Dine supersymmetric baryogenesis \citep{ad-js,ad-mk-nk} were shown to be able to match the observed properties of GW150914 \cite{2016JCAP...11..036B}.

\section{Low spins of BH in GW150914 and LTV151012 events}

In the framework of general relativity, a BH is fully characterized by its mass $M$ and dimensionless 
angular momentum $a=J/M$ (in geometrical units $G=c=1$) (the possible BH electric charge is negligible in real astrophysical conditions). 
The LIGO observations enable measurements of both masses of the coalescing BH components, $M_1$ and $M_2$, and the chirp mass that determines the strength of gravitational wave signal ${\cal M}=(M_1M_2)^{3/5}/M^{1/5}$. From the analysis of waveforms at the inspiral stage, individual BH spins before the merging are poorly constrained, but their
mass-weighted total angular momentum parallel to the orbital angular momentum, $\chi_{eff}$, can be estimated with good accuracy 
\citep{2016PhRvX...6d1015A}\footnote{The parameter $\chi_{eff}=(M_1\chi_1+M_2\chi_2)/M$, where $\chi_i=a_i\cos\theta_i$ with $\theta_i$ being the angle between the angular momentum of the i-th BH and orbital angular momentum of the binary system.}. The O1 LIGO detections suggest that the most massive GW150914 and (less certain) LTV151012 have very low $\chi_{eff}\simeq 0$.

\begin{figure*}
\begin{center}
 \includegraphics[width=3.4in]{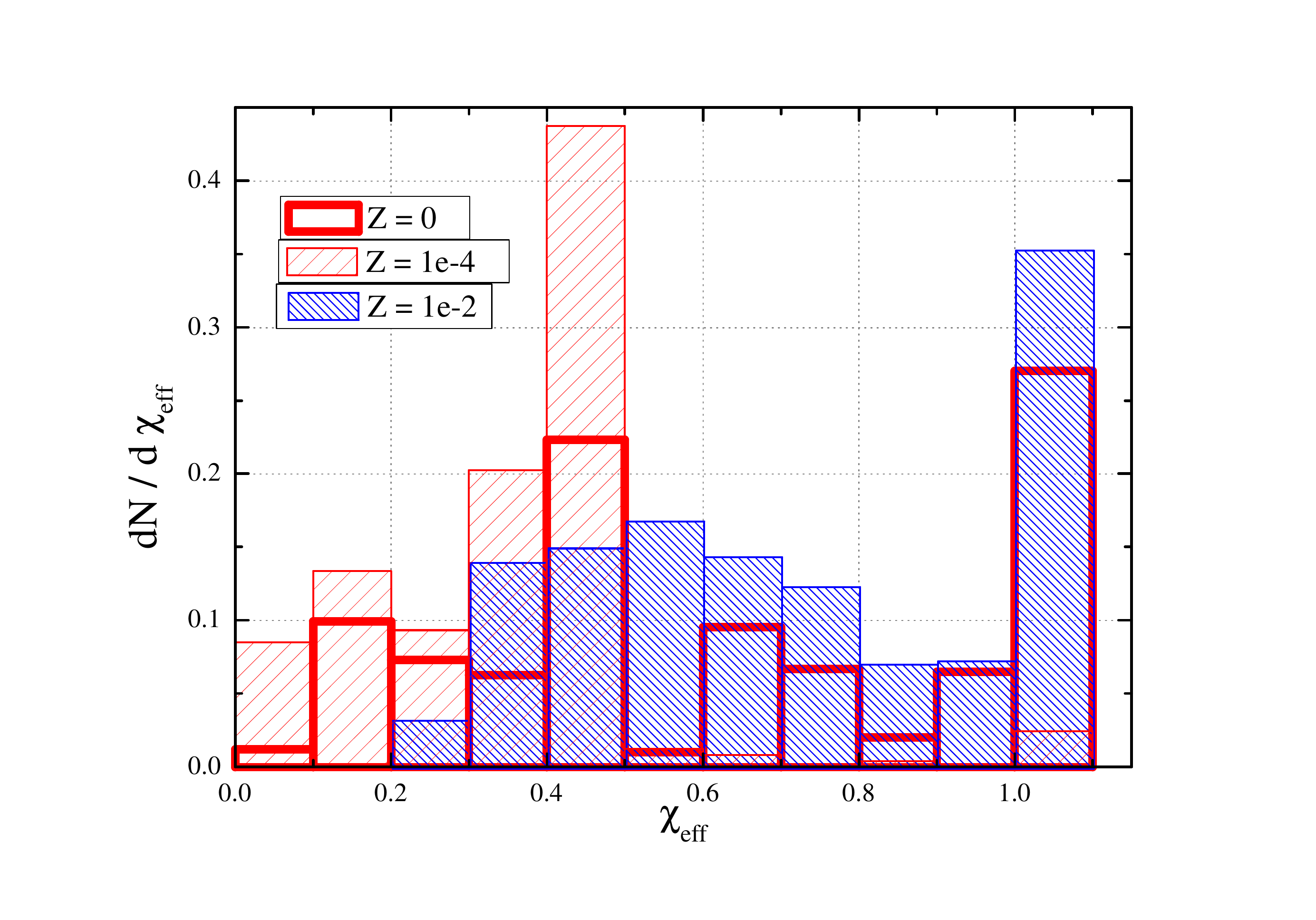} 
 \caption{Distribution of the effective binary BH spin parameter $\chi_{eff}$ before the merging for different stellar metallicities.}
   \label{f:chi}
\end{center}
\end{figure*}

\begin{figure*}
 \includegraphics[width=0.85\textwidth]{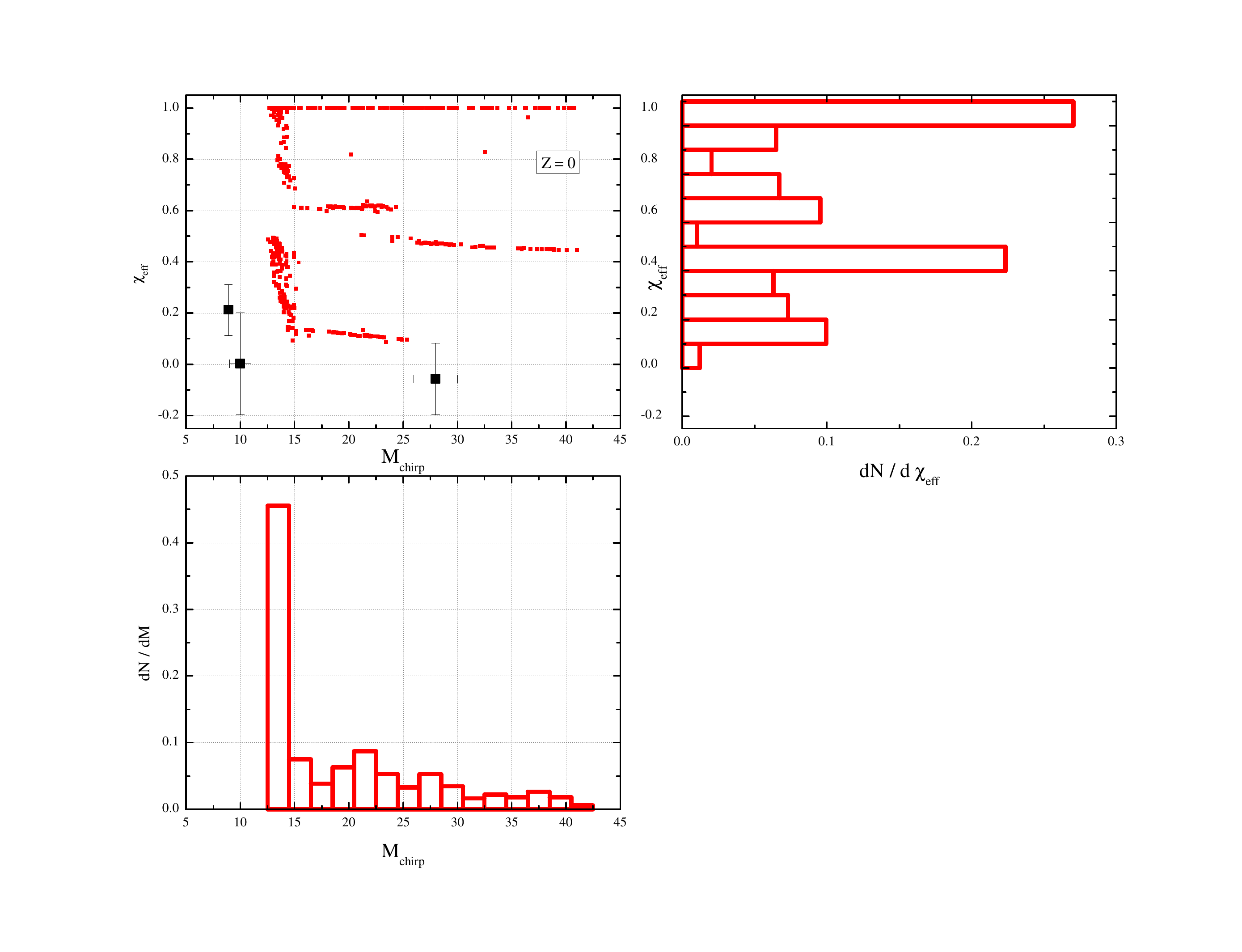} 
\vfill
 \vspace*{-1.0 cm}
\includegraphics[width=0.85\textwidth]{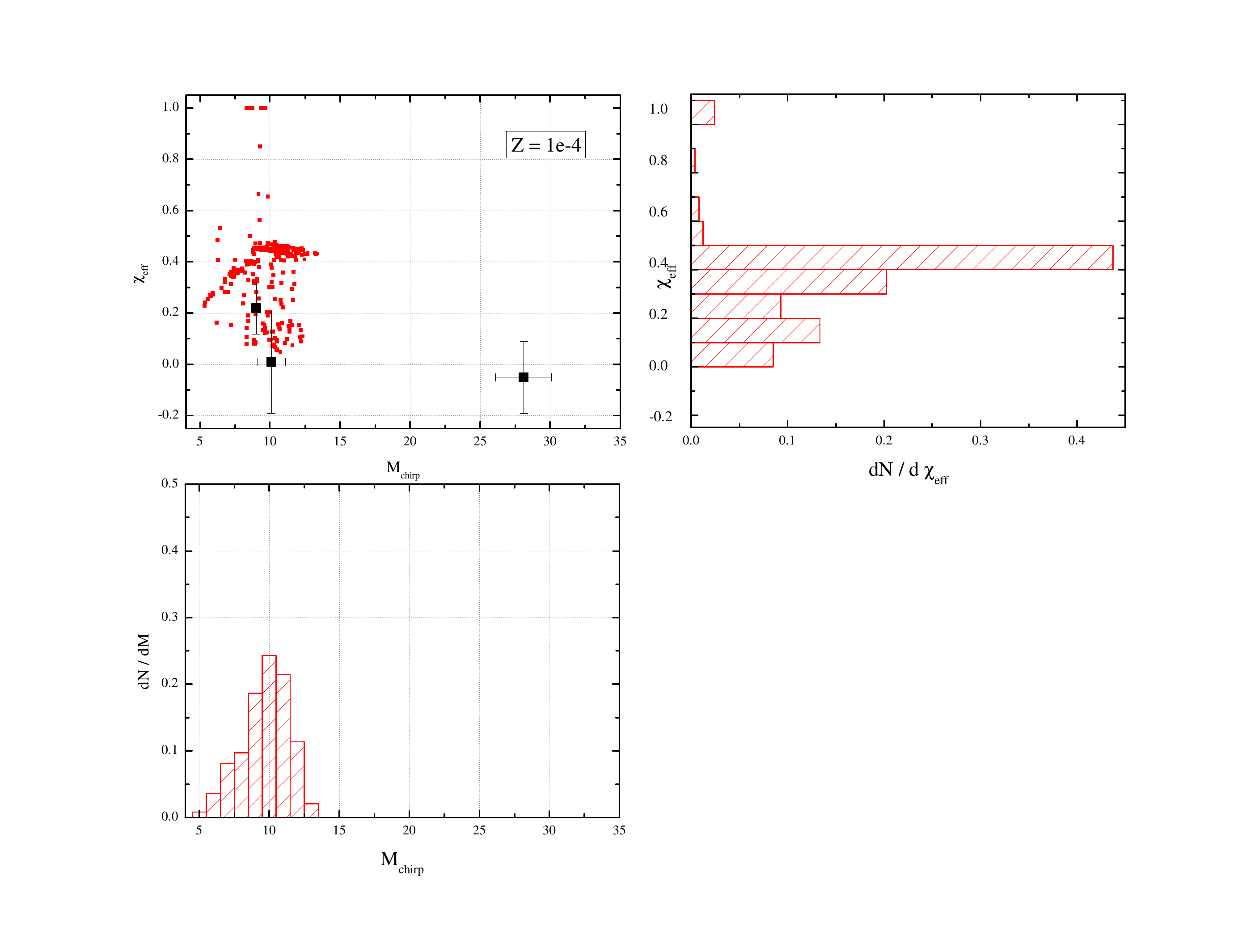} 
\vfill
\vspace*{-1.0 cm}
\includegraphics[width=0.85\textwidth]{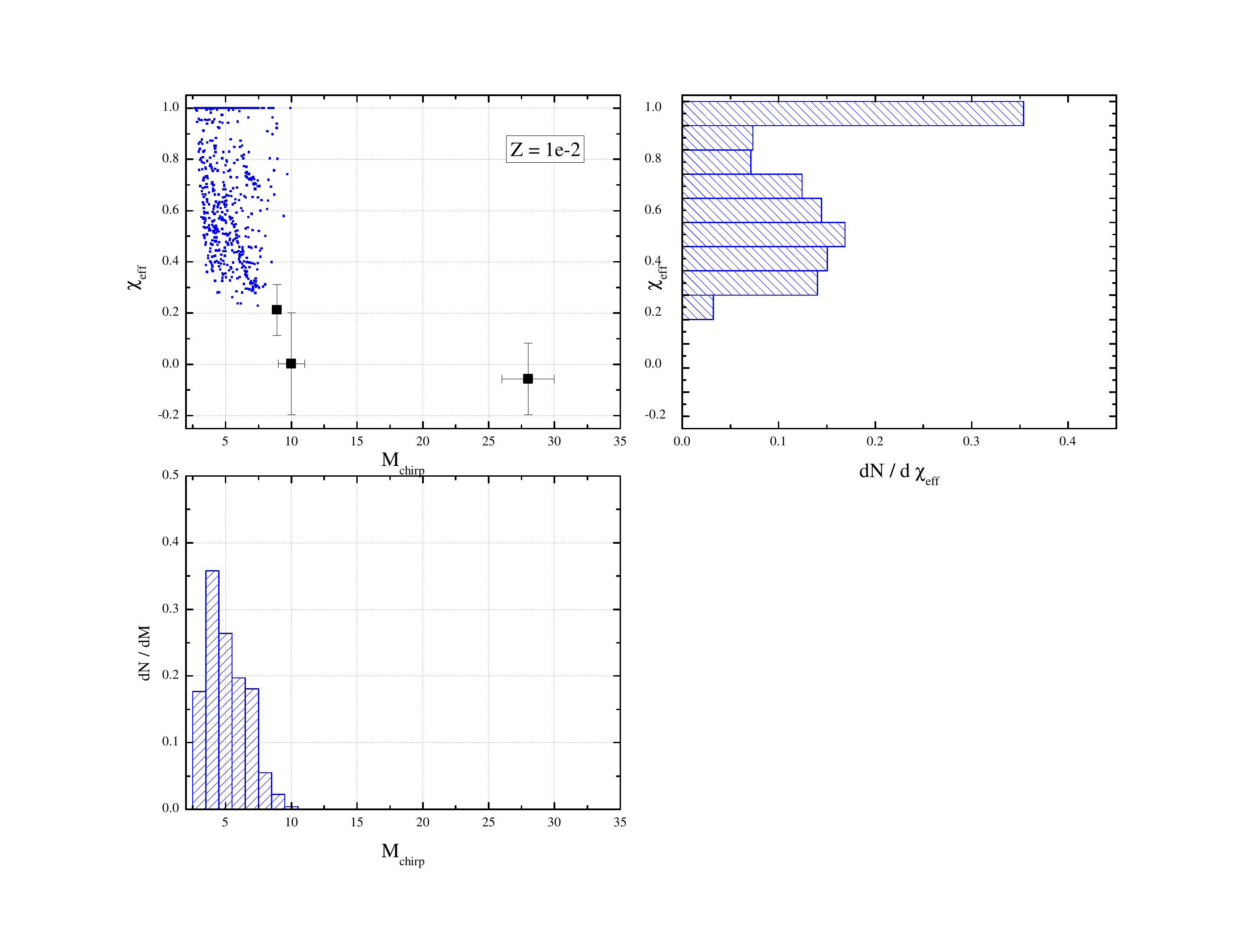} 
 \caption{${\cal M}-\chi_{eff}$ plane for different stellar metallicities. Filled squares show the observed BH-BH systems \citep{2016PhRvX...6d1015A},
GW150914, LTV151012 and GW151226, in order of decreasing chirp mass.}
   \label{f:Mchi}
\end{figure*}

This observational fact has important evolutionary implications (see \cite{2016MNRAS.462..844K,2017arXiv170203952H}). It suggests a very slow rotation of BH progenitors, which by itself strongly constraints, for example, chemically homogeneous pathways mentioned above in which tidally induced rotation of the close binary  
components plays the key role. Massive stars are observed to be rapid rotators. No significant angular momentum loss is expected during
their evolution with low mass loss rate by stellar wind and at the pre-collapse stage as required to produce massive BHs \citep{2015MNRAS.451.4086S}. 
Note that low effective spin values can imply either small intrinsic BH spins $a\sim 0$, or unusual orientations of BH spins with respect to the 
orbital angular momentum at the inspiral stage. The last case can well be reconciled with the dynamical formation scenario \citep{2016PhRvD..93h4029R}, where the BH spins are not expected to be correlated with the orbital angular momentum. In the PBH scenario, BH spins must be zero as there are no vorticity 
in primordial cosmological perturbations.

Therefore, the mass-spin distribution of BHs can serve as a sensitive tool to discriminate between different 
astrophysical formation channels of coalescing massive binary BHs. To estimate the spin distribution of BH remnants in binaries, 
it is necessary to know how to treat the spin evolution of the stellar core, which is ill-understood and strongly model-dependent. 
One possible approach is to match theoretical predictions of the core rotation with observed period distribution of the young neutron stars observed as radio pulsars \citep{2016MNRAS.463.1642P}. Initially, a star is assumed to rotate rigidly, but after the main sequence the star can be separated in
two parts -- the core and the envelope, with some effective coupling between these two parts. The coupling between the core and envelope rotation can be mediated by magnetic forces, internal gravity waves (see \cite{2015ApJ...810..101F} and J.Fuller's talk at this conference), etc.
The validity of such an approach was checked by direct MESA calculations of the rotational evolution of a 15~$M_\odot$ star \citep{2016MNRAS.463.1642P}.
It was found that the observed period distribution of young pulsars can be reproduced if the effective coupling time between the core and envelope is $\tau_c=5\times 10^5$~years (see Fig. 1 in \cite{2016MNRAS.463.1642P}). Below we shall assume that this time is also applicable to the evolution of very massive stars leaving behind BH remnants.

Each angular momentum of the main-sequence components of the initial binary is assumed to be arbitrarily distributed in space, its absolute 
value being connected to the initial stellar mass using the empirical relation between the equatorial rotation velocity of a star with its mass
$v_{rot}=330M_0^{3.3}/(15+M_0^{3.45})$~km~s$^{-1}$ (here $M_0$ is in solar units). It was assumed that the rotation of the stellar envelope gets synchronized with the orbital motion with the characteristic synchronization time $t_{sync}$, and the process of tidal synchronization was treated as in the BSE code \citep{2002MNRAS.329..897H}. Due to intrinsic misalignment of the spin vectors of the stars with the binary orbital angular momentum $\hat L$, we separately treated the core-envelope coupling for the spin components parallel and perpendicular to $\hat L$. On evolutionary stages prior to the compact remnant formation, for each binary component we assumed that due to tidal interactions the parallel component of the stellar envelope spin $J_{||}$ gets synchronized with the orbital motion on  the characteristic time $t_{sync}$, while the normal spin component $J_{\perp}$ of the stellar envelope decreases due to the tidal interaction in the binary system the on 
the same characteristic time scale, which leads to the secular evolution of the spin-orbit misalignment. The parallel component of the 
envelope spin also evolves due to the core-envelope interaction with the characteristic time $\tau_c$. These processes were added to 
the updated BSE population synthesis code. 

\begin{figure*}
\begin{center}
 \includegraphics[width=0.31\textwidth]{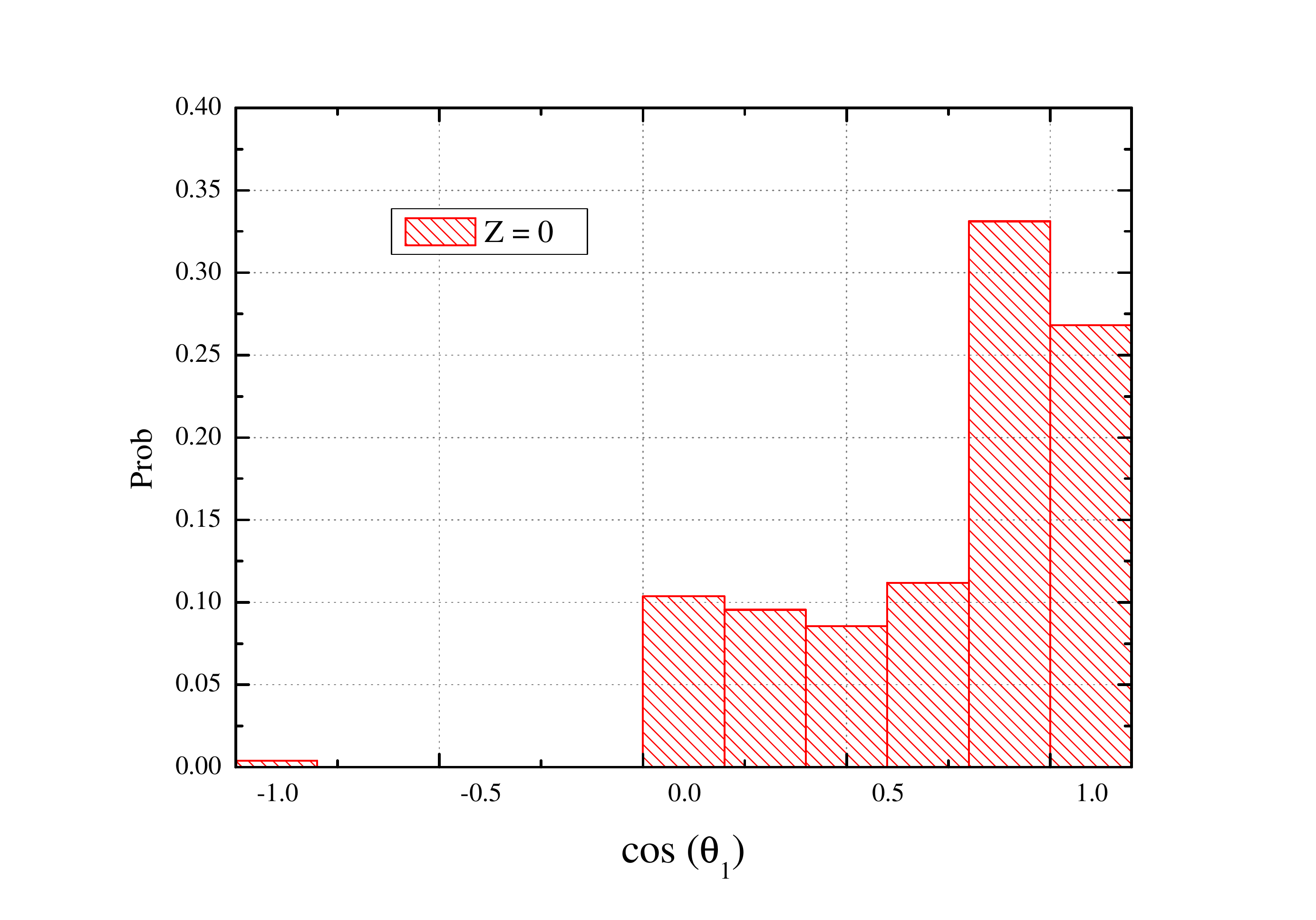} 
\hfill
\includegraphics[width=0.31\textwidth]{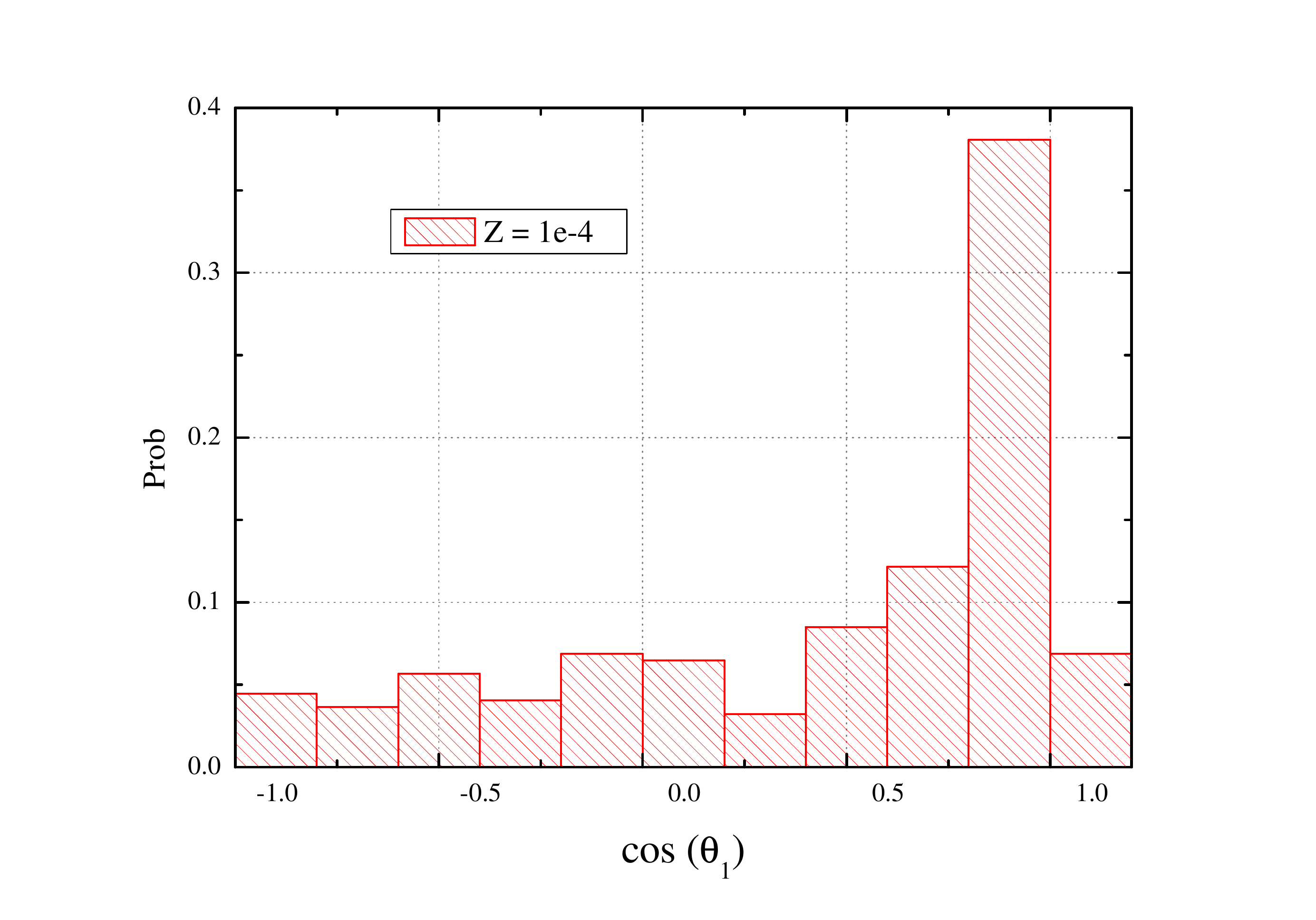} 
\hfill
\includegraphics[width=0.31\textwidth]{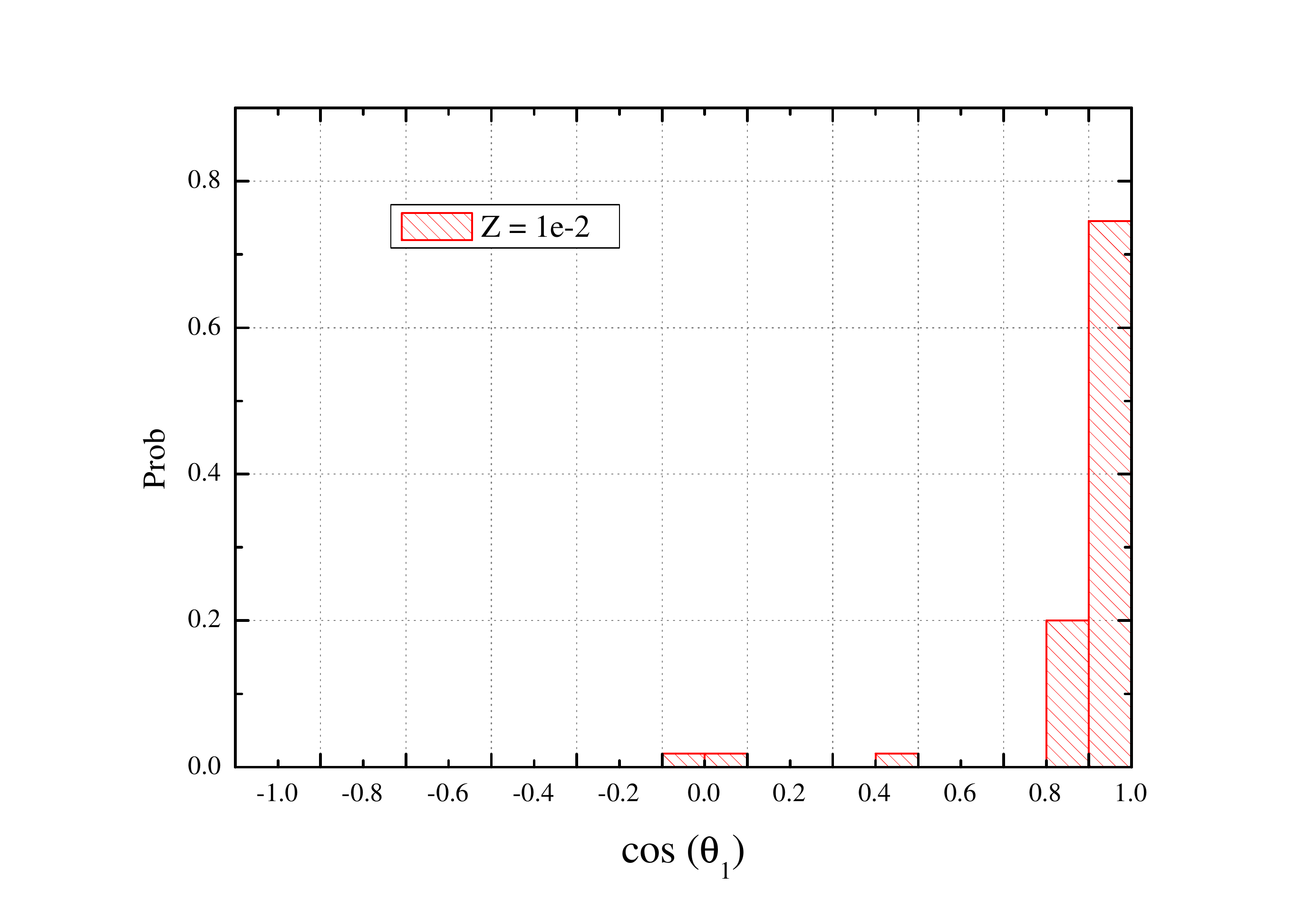} 
\vfill
\includegraphics[width=0.31\textwidth]{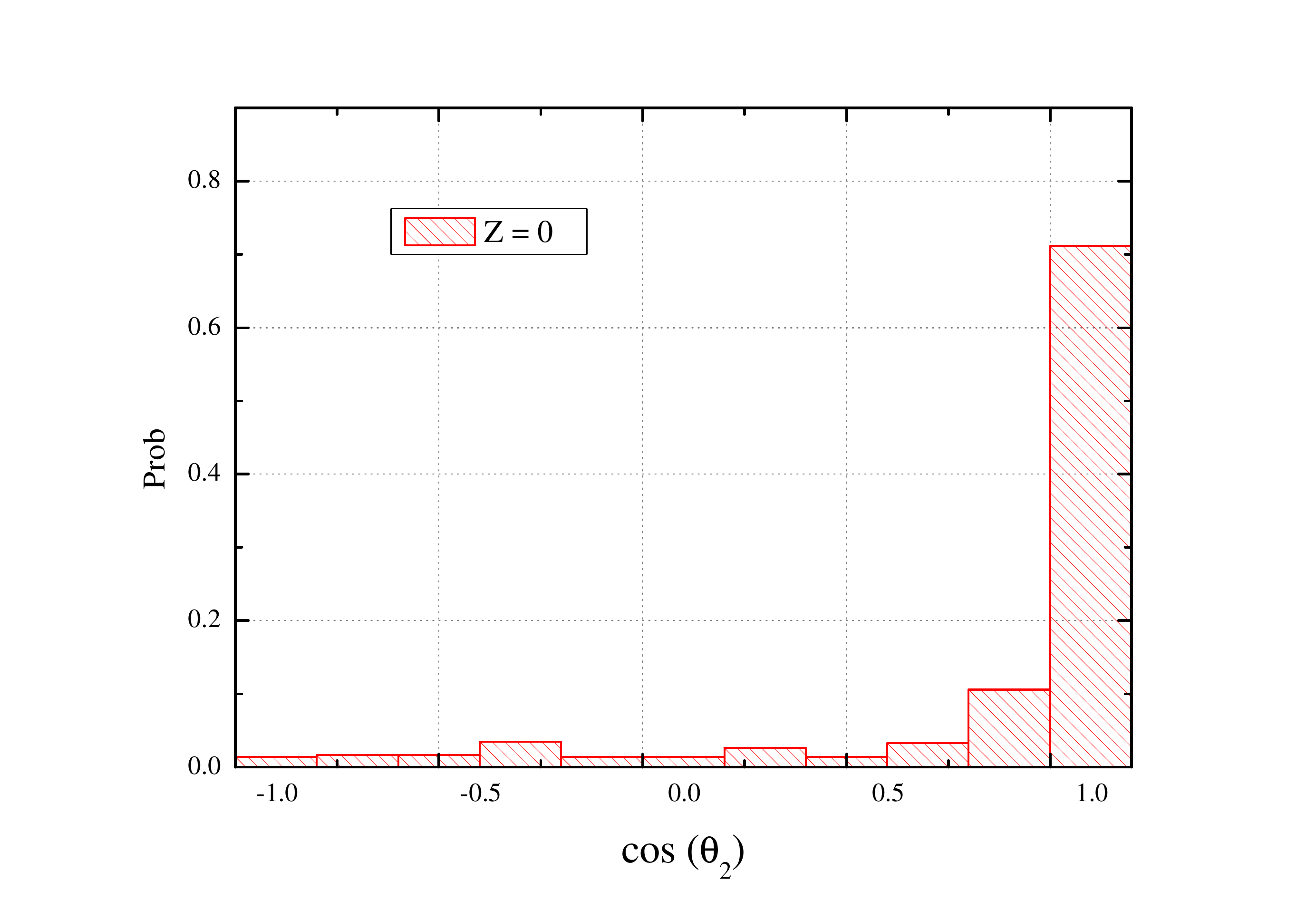} 
\hfill
\includegraphics[width=0.31\textwidth]{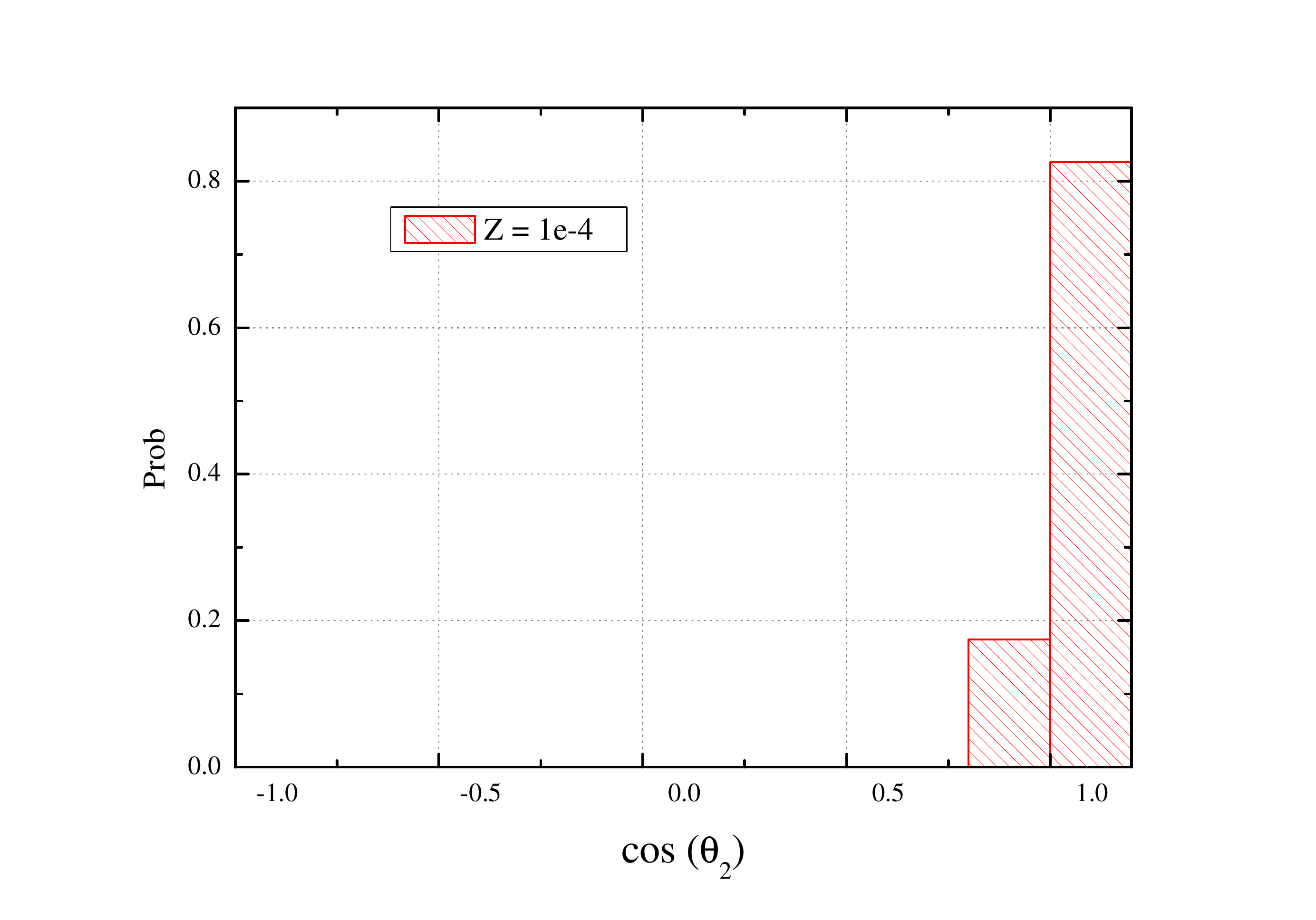} 
\hfill
\includegraphics[width=0.31\textwidth]{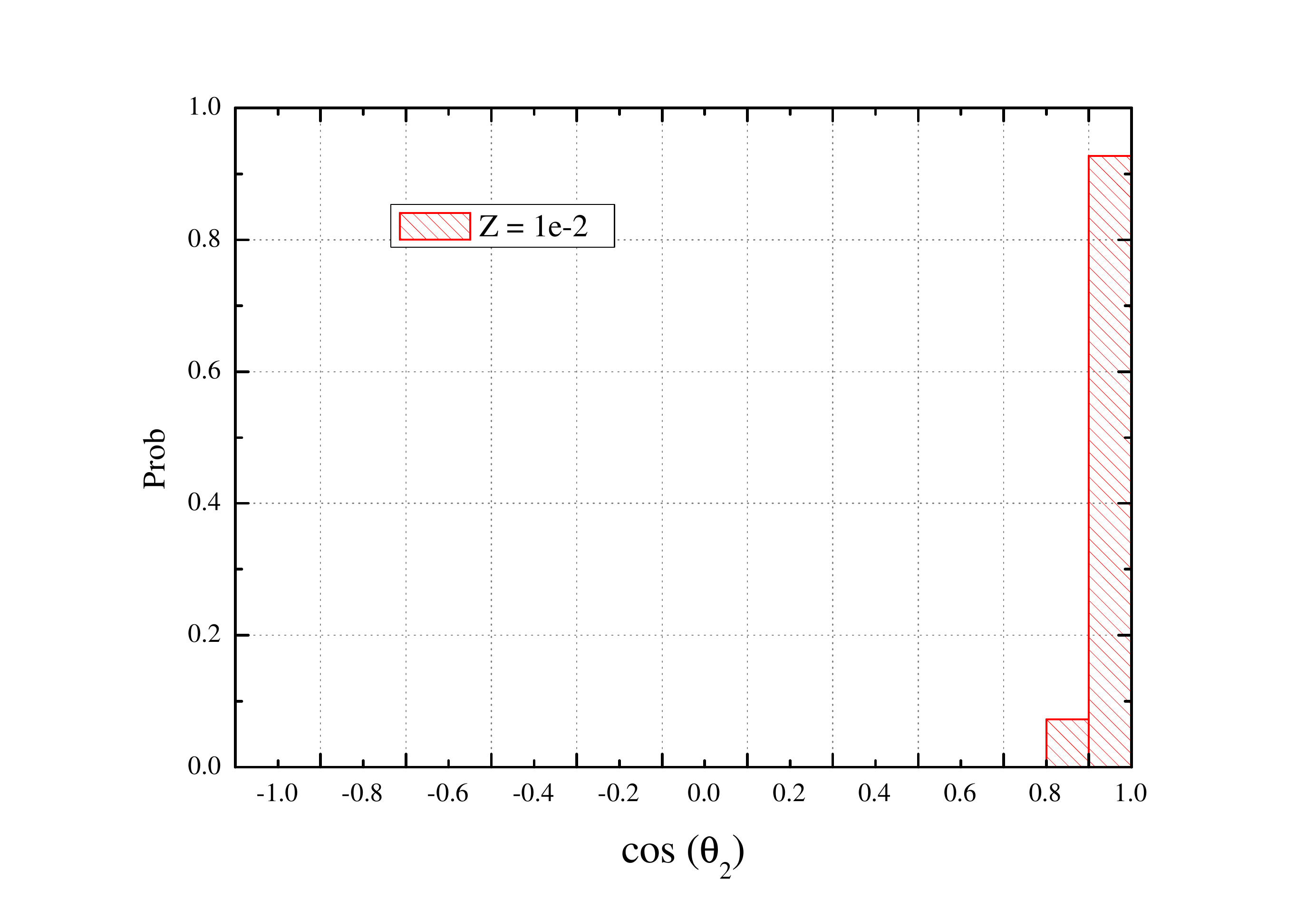} 
 \caption{BH spin misalignments  with orbital angular momentum (in terms of $\cos\theta$) in coalescing binary BHs for different stellar metallicities.
Top and bottom row correspond to $M_1$ and $M_2$, respectively.}
   \label{f:costheta}
\end{center}
\end{figure*}

With these additions, the population synthesis of typically 100000 binaries per run has been carried out for different parameters of binary evolution (the common envelope stage efficiency $\alpha_{CE}$, stellar metallicities etc.). 
No generic BH kick was assumed. The results of calculations of BH spin distributions for different stellar metallicites and for the standard CE efficiency parameter $\alpha_{CE}=1$ are shown in Fig.  \ref{f:chi}. Evolution of zero-metallicity (primordial Population III) stars was parametrized as in \cite{Kinugawa2014}. 
Fig. \ref{f:Mchi} shows the plot of coalescing binary BHs on the ${\cal M}-\chi_{eff}$ plane for different metallicities.
BH spin misalignments with orbital angular momentum in coalescing binary BHs for different stellar metallicities are presented in Fig. \ref{f:costheta}.

A detailed analysis of these simulations will be published elsewhere (Postnov \& Kuranov, in preparation), but the main conclusions 
can be drawn from Figs. 1-3. It is seen (expectedly) from Fig. 1 that the effective spin $\chi_{eff}$ of binary BH from field massive stars (the standard formation scenario) is distributed in a wide range, but the ${\cal M}-\chi_{eff}$ plot (Fig. 2) suggests that large chirp masses 
can hardly have $\chi_{eff}\simeq 0$. This (model) result can signal potential difficulty in explaining the most massive merging BH binaries by this formation channel only. Fig. 3 suggests that even in the absence of BH kicks, which were assumed in the present calculations, the BH spin
misalignments can be quite high even for field binaries.

\section{Conclusions}

Presently, there are different astrophysical pathways of producing massive binary BHs that merge in the Hubble time.
They can be formed from low-metallicity massive field stars, primordial Pop III remnants, can be results of dynamical evolution in dense 
stellar clusters or even primordial black holes. It is not excluded that all channels contribute to the observed 
binary BH population. For example, the discovery of very massive Schwarzschild BHs would be difficult to reconcile with the
standard massive binary evolution, but can be naturally explained in the PBH scenario \citep{2016JCAP...11..036B}. 

As of the time of writing, 
another two event candidates were reported by the LIGO collaboration from the analysis of 12 days of joint operation of two LIGO interferometers during O2 run (see 
\verb+http://ligo.org/news/index.php#O2Jan2017update+). With the current LIGO sensitivity, the detection horizon of binary BH with masses around 30 $M_\odot$ 
reaches 700 Mpc. So far the statistics of binary BH merging rate as a function of BH mass as inferred from three
reported LIGO O1 events is consistent with a power-law dependence, $dR/dM\sim M^{-2.5}$ \citep{2017arXiv170203952H}, which does not contradict the general power-law behavior of the stellar mass function. Clearly, more statistics of BH masses and spins inferred from 
binary BH mergings is required to distinguish between the possible binary BH populations which can exist in the Universe. 

\vskip\baselineskip
\textbf{Acknowledgements.} KP acknowledges the support from RSF grant 16-12-10519.


\end{document}